\documentclass[aps,prb,twocolumn,superscriptaddress,floatfix]{revtex4}
\usepackage{graphicx}
\usepackage{epsfig}
\usepackage{amsmath}
\usepackage{amssymb}
\usepackage{subfigure}
\usepackage{relsize}
\usepackage{mathptmx}
\usepackage{microtype}

\begin{document}


\title{Common origin of exotic properties in ceramic and hybrid negative thermal expansion materials}

\author{Hong Fang}
\affiliation{Department of Earth Sciences, University of Cambridge, Downing Street, Cambridge, CB2 3EQ, U.K.}

\author{Martin T. Dove}
\email{martin.dove@qmul.ac.uk}
\affiliation{Department of Earth Sciences, University of Cambridge, Downing Street, Cambridge, CB2 3EQ, U.K.}
\affiliation{Centre for Condensed Matter and Materials Physics, School of Physics and Astronomy, Queen Mary University of London, Mile End Road, London, E1 4NS, U.K.}
\affiliation{Materials Research Institute, Queen Mary University of London, Mile End Road, London, E1 4NS, U.K.}

\author{Anthony E. Phillips}
\affiliation{Centre for Condensed Matter and Materials Physics, School of Physics and Astronomy, Queen Mary University of London, Mile End Road, London, E1 4NS, U.K.}
\affiliation{Materials Research Institute, Queen Mary University of London, Mile End Road, London, E1 4NS, U.K.}

\date{\today}
\begin{abstract}
By considering the lattice dynamics between bond strain and transverse vibration, we establish a common model for ceramic and hybrid materials that exhibit negative thermal expansion (NTE). We show that pressure-induced softening, increase in NTE with pressure, and variation of NTE with temperature all arise naturally from the same mechanical model that gives rise to the NTE itself.
\end{abstract}


\maketitle

\section{Introduction}

Many ceramic and hybrid metal-organic framework materials show negative thermal expansion (NTE): they \textit{contract} instead of expanding on heating~\cite{Barrera2005,Miller2005,Lind2012,Romao2013}. Their structures are invariably characterised as a network of polyhedral groups of atoms that are connected through sharing of corner atoms or by shared ligands. Empirically, NTE materials tend to show pressure-induced softening, pressure enhancement of NTE, and the reduction of NTE on heating. But such effects have only been investigated in a small number of materials \cite{Pantea 2006,Chapman 2005,Chapman 2007,Fangexp 2013}, and as yet there is no general framework for understanding the whole suite of properties together.

Very recently we have seen a new emphasis on the role of anharmonic interactions in influencing the properties of NTE materials. Recently experimental and theoretical work on some simple materials, including ScF$_3$~\cite{Li 2011}, Cu$_2$O~\cite{Gupta2014} and Ag$_2$O~\cite{Lan2014}, has suggested that the extent of NTE at high temperature can only be understood on the basis of an important role for intrinsic phonon anharmonic interactions.

In this paper we present a study of simplified models of NTE materials (Section II) in which the Hamiltonians are chosen to reflect the physical picture generally accepted as responsible for NTE in framework materials. In Section III we use these models in the quasi-harmonic approximation to demonstrate that NTE, pressure-enhanced NTE, and pressure-induced softening naturally emerge together. Finally, in Section IV, we show how including anharmonic interactions in these models leads to structural warm hardening---something that has only previously been seen in laser-excited warm-dense matter~\cite{Ernstorfer 2009} and Sc$_{1-x}$Y$_x$F$_3$ ($x\leq0.25$) solid solution~\cite{Morelock 2013}---as well as to the transition from NTE to positive thermal expansion and the disappearing of the pressure-induced softening at high temperatures.

\section{Simple models of NTE materials}

We present here two simple one-dimensional models designed to capture the essential physics and chemistry of NTE in ceramic and metal-organic framework (MOF) materials.
These materials consist of metal atoms, which we denote by \textbf{M}, connected by one or more linking atoms that we denote by \textbf{X}. Their NTE behaviour is typically explained in terms of the transverse vibrations of the \textbf{X} atoms, which, because of tight bonds to their \textbf{M} neighbours, pull the \textbf{M} atoms towards one another~\cite{Barrera2005,Miller2005,Lind2012,Romao2013,Welche1998,Heine1999,Simon2001,He2010}. Moreover, it is generally most favourable for the coordination polyhedron surrounding each \textbf{M} atom to rotate as a rigid unit~\cite{Barrera2005,Miller2005,Lind2012,Romao2013,Welche1998,Heine1999,Giddy 1993} so that most of the flexing of the framework occurs about the \textbf{X} atoms. To capture this physics, our model Hamiltonian must contain three components: strong penalties for distorting both (1) nearest-neighbour bonds and (2) coordination polyhedra, as well as (3) a weaker penalty for flexing about other points of the chain. We will construct two such Hamiltonians: one to describe ceramics, in which neighbouring metal atoms are connected by a single \textbf{X} atom (typically representing oxygen), and one to describe MOFs, in which neighbouring metal atoms are connected by a pair of \textbf{X} atoms representing a rigid molecular linker. By design, these models include the minimum possible description of the mechanism for NTE in broad classes of ceramic and MOF materials, with the intention to demonstrate that the concomitantly observed physical properties do not arise by coincidence, but rather emerge naturally from the same Hamiltonian.


Consider first ceramic materials; examples with NTE include ZrW$_2$O$_8$ \cite{Mary1996,Evans1996}, ZrV$_2$O$_7$ \cite{Korthuis 1995}, Sc$_2$W$_3$O$_{12}$ \cite{Evans1997,Mary1999}, Cu$_2$O \cite{Sanson2006,Rodriguez2009}, ScF$_3$ \cite{Greve 2010,Li 2011}, quartz and a number of zeolites \cite{Attfield 1998,Lightfoot 2001,Fangzeolite 2013}.  \begin{center} \includegraphics[scale=0.75]{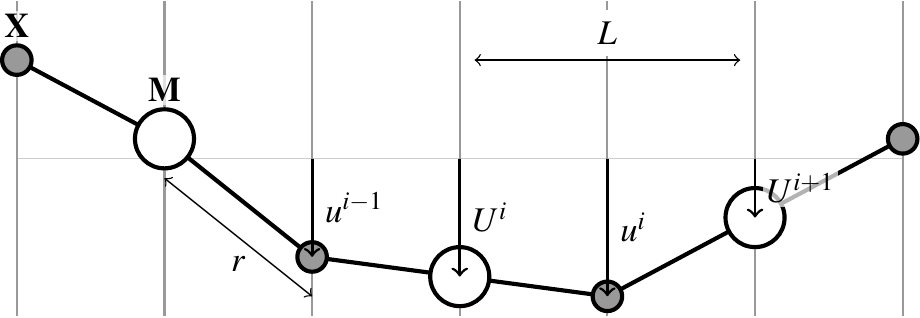} \end{center}

We model these materials as a chain of equally-spaced, alternating \textbf{M} and \textbf{X} atoms, with bond length $r$ and lattice parameter $L$. For simplicity we allow the atoms to move only transverse to the chain, with displacements of the \textbf{X} and \textbf{M} atoms denoted respectively by $u^i$ and $U^i$: $i$ indexes the unit cell. The lattice strain is denoted by $e$, with $L = L_0(1+e)$ and $L_0$ is the equilibrium lattice parameter when the atoms are at rest.

The three terms in the Hamiltonian corresponding to the discussion above are as follows. The first term describes the energy cost of changing the length of the \textbf{M}--\textbf{X} bonds, $r$, due to both atomic displacements and elastic strain:
\begin{equation}\label{potential_1}
V_1  = \frac{k}{2} \sum\limits_{i,j} \left[ \sqrt {r_0 ^2 \left( 1 + e \right)^2  + \left(  u^{i+j} -U^i \right)^2 }- r_0  \right]^2
\end{equation}
with $j\in\{-1,0\}$ and $r_0=L_0/2$.
The second energy term reflects the energy cost of distorting an \textbf{MX}$_2$ polyhedron through bending the \textbf{X--M--X} bond:
\begin{equation}\label{potential_2}
V_2  =  \frac{K}{2} \sum_i{ \left( {u ^i  + u ^{i - 1}  - 2U ^i } \right)^2  }.
\end{equation}
Finally, the third term provides weaker resistance against \emph{any} relative displacements of neighbouring atoms:
\begin{equation}\label{potential_3}
V_3  = \frac{J}{2} \sum_i  \left[ (   u^{i-1} - U^i )^2  + ( u^i - U^i )^2 \right].
\end{equation}

Picturing the \textbf{MX}$_2$ unit as mechanically rigid, it is instructive to analyse this system in terms of its rigid unit modes (RUMs). Following the analysis of Ref.~\onlinecite{Giddy 1993}, the \textbf{MX}$_2$ unit will have two degrees of freedom---one rotation and one translation---and one constraint per unit, meaning that there will be one RUM at each wave vector. The system will have two branches in its phonon dispersion curve, with the RUM corresponding to the acoustic mode. The first two energy terms will give exactly zero frequency for each RUM; the third term provides some resistance against these distortions.

We now adapt this to form an analogous model representing a hybrid MOF system, such as Zn(CN)$_2$, Cd(CN)$_2$ and related materials \cite{Goodwin2005,Goodwin2005b,Chapman2006,Phillips2008} as well as a wide range of hybrid MOFs \cite{Rowsell2005,Han2007,Lock 2010}.
\begin{center}
\includegraphics[scale=0.75]{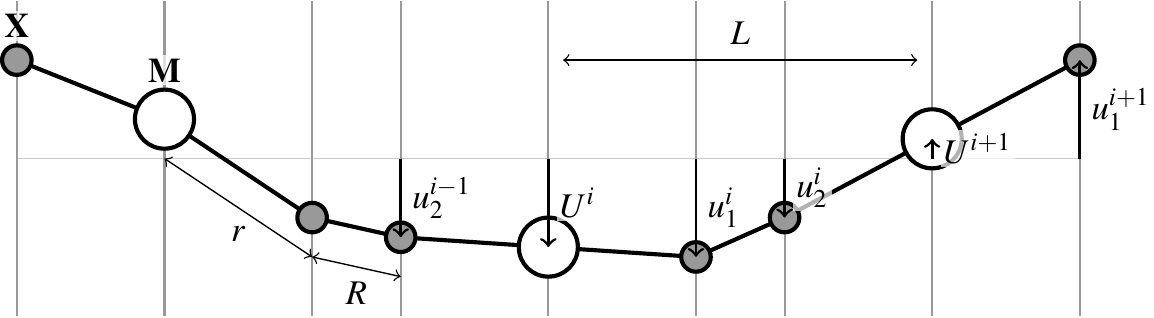}
\end{center}

In this model the \textbf{X}--\textbf{X} linkage represents a molecular ligand---such as CN in Zn(CN)$_2$~\cite{Fangmd 2013} or C$_6$H$_4$(COOH)$_2$ in \mbox{MOF-5}~\cite{Lock 2010}---with length $R$ (equilibrium length $R_0$). Our three potential energy terms are modified to
\begin{multline}
  V_1^\prime  = \frac{k}{2} \sum\limits_{i} \left[ \sqrt {r_0 ^2 \left( 1 + e \right)^2  + \left(  u^{i}_1 -U^i \right)^2 }- r_0  \right]^2 + \\
  \frac{k}{2} \sum\limits_{i} \left[ \sqrt {r_0 ^2 \left( 1 + e \right)^2  + \left(  u^{i-1}_2 -U^i \right)^2 }- r_0  \right]^2 + \\
  \frac{k^\prime}{2}\sum\limits_{i} {\left[ {\sqrt {R_{0} ^2 \left( {1 + e^\prime } \right)^2  + \left( {u_\mathrm{2} ^{i}  - u_\mathrm{1} ^{i} } \right)}  - R_{0} } \right]^2 }
\end{multline}
\begin{equation}
V_2'  =  \frac{K}{2} \sum_i{ \left( {u_1 ^i  + u_2 ^{i - 1}  - 2U ^i } \right)^2  }
\end{equation}
\begin{equation}
V_3'  = \frac{J}{2} \sum_i  \left[ (   u_2^{i-1} - U^i )^2  + ( u_1^i - U^i )^2 \right] + \\
\frac{J^\prime}{2}\sum\limits_i{\left( {u_\mathrm{2} ^i  - u_\mathrm{1} ^{i} } \right)^2}
\end{equation}
where $e$ and $e^\prime$ represent strains in $r$ and $R$, respectively, and we set $k'/k = r_0/R_0$ to make the strain homogeneous.

The long-wavelength solutions to the dynamical matrix of the ceramic model when the two masses are set equal to 1 are, to first order in $e$,
\begin{equation}\label{freq_longwave}
  \omega ^2|_{q\rightarrow 0} = \left\{ {\begin{array}{*{20}c}
      {\left(J+k e\right)\left(q^2 /4\right)}  \hfill\quad\text{(Acoustic)} \\
   {}  \\
   {K\left( {8 - q^2} \right)+\left(J  + k e\right)\left( {4 - q^2/4} \right)} \hfill\quad\text{(Optic)} \\
\end{array}} \right.
\end{equation}
where $q$ is the wave vector divided by $a^*=2\pi/L$.
The solutions for the zone boundary wave vector are
\begin{equation}\label{freq_boundary}
  \omega ^2|_{q = 1/2} = \left\{ {\begin{array}{*{20}c}
        {2J + 2ke{\rm{   }}} \hfill\quad\text{(Acoustic)} \\
 {}  \\
   {4K + 2J + 2ke} \hfill\quad\text{(Optic).} \\
\end{array}} \right.
\end{equation}
As anticipated, the acoustic mode at this wave vector is the RUM, with eigenvector $[U,u]=[0,1]$, while the optic mode has eigenvector $[1,0]$.

It is clear from these two sets of solutions---again as anticipated above---that setting $J=0$, and hence $V_3=0$, will cause the RUMs to have zero frequency when $e = 0$. Moreover, setting $K=0$ and hence $V_2=0$ gives equal frequencies for the two zone-boundary modes: without $V_2$ we only have a simple monatomic chain.

\begin{figure}[t]
\begin{center}
\includegraphics[width=7.2cm]{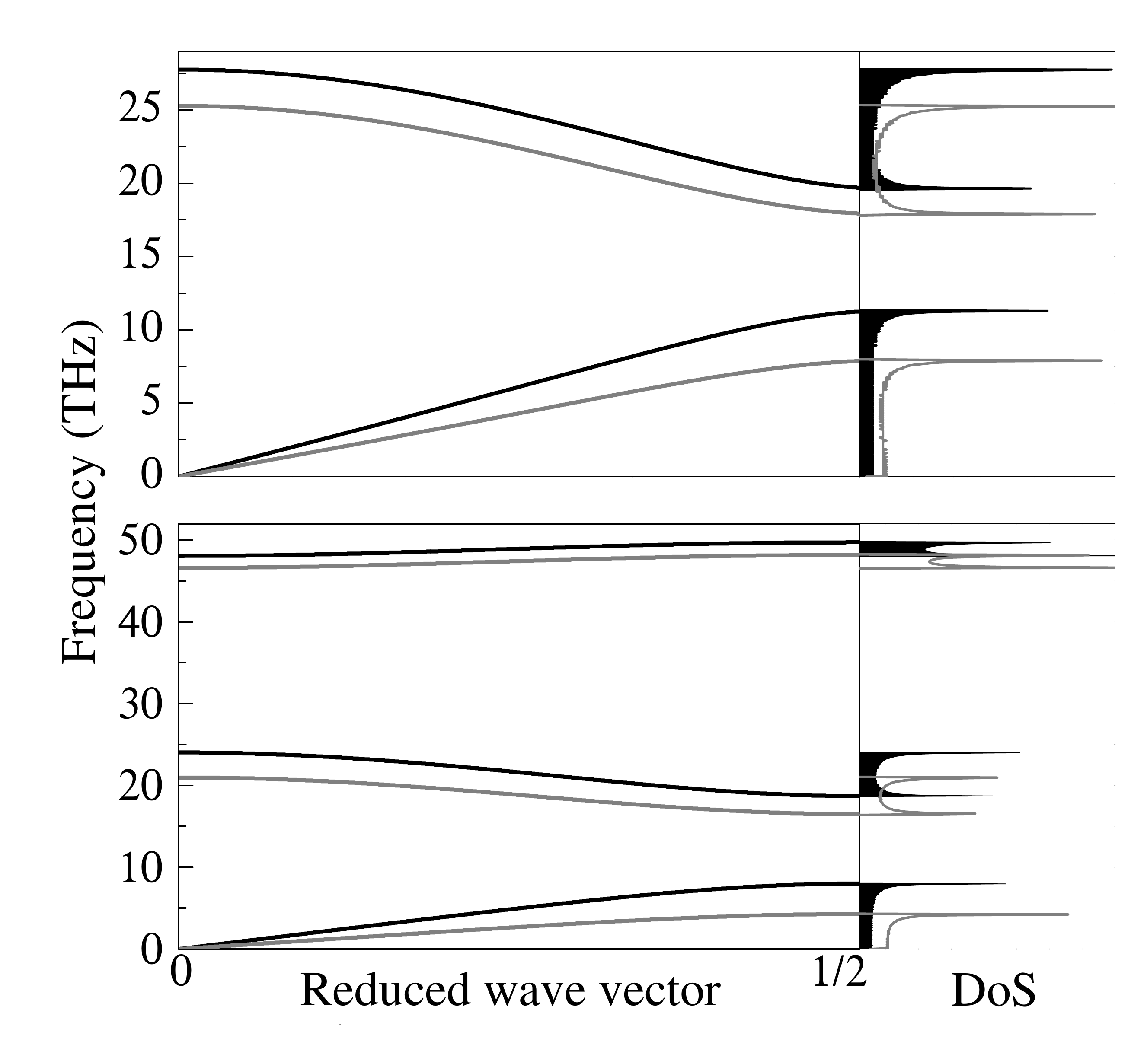}
\end{center}
\caption{\label{phonons} The phonon dispersion curves (left) and densities of states (right) for the ceramic (top) and hybrid (bottom) models. Results are shown for two values of the lattice parameter, corresponding to the equilibrium values at zero pressure (black) and a compressive strain of $e = -0.02$ (grey).}
\end{figure}

The phonon dispersion curves and density of states for both models for a particular set of parameters (see `Methods') and for two fixed strains are shown in Figure \ref{phonons}. For all branches (indexed by $s$) the frequency values \textit{decrease} on compression (negative $e$), giving negative values of the mode Gr\"{u}neisen parameter $\gamma_{s,q} = -\partial \ln \omega(s,q) / \partial \ln L$.

In the hybrid model, the optic modes at $q=0$ correspond to neighbouring \textbf{X}--\textbf{M}--\textbf{X} rotating in the opposite (lower-frequency) and the same (higher-frequency) sense. The value of $J^\prime$ determines the splitting of these frequencies.

Working with a model stripped down to interactions along a one-dimensional chain and with one transverse degree of freedom per atom, whilst maintaining a three-dimensional nature of the interatomic interactions, makes for an easier and more transparent analysis. In particular, we reduce the number of irrelevant degrees of freedom, and we reduce the number of parameters required to define a realistic model. There are two main effects from working with a one-dimensional topology. The first is on the value of the pressure-derivative of the bulk modulus at zero temperature. A three-dimensional harmonic potential $\Phi  = (k/2)(V^{1/3} - V_0 ^{1/3})^2$ gives $B_0^\prime=1$ and, for a general interatomic potential $\Phi = -ar^{-m}+br^{-n}$, $B_0^\prime=(m+n+6)/3$. A one-dimensional topology with a harmonic potential gives $B_0^\prime=-1$. Second is the effect on the phonon density of states, which will have a constant value at low $\omega$ rather than varying as $\omega^2$. This will lead, for example, to the heat capacity at low temperature varying as $T$ rather than $T^3$ at low temperature. However, all the important points of the paper are related to the form of the vibrations and are not sensitive to the effect of dimensionality on thermodynamics.

\begin{figure}[t]
\begin{center}
\includegraphics[width=8.0cm]{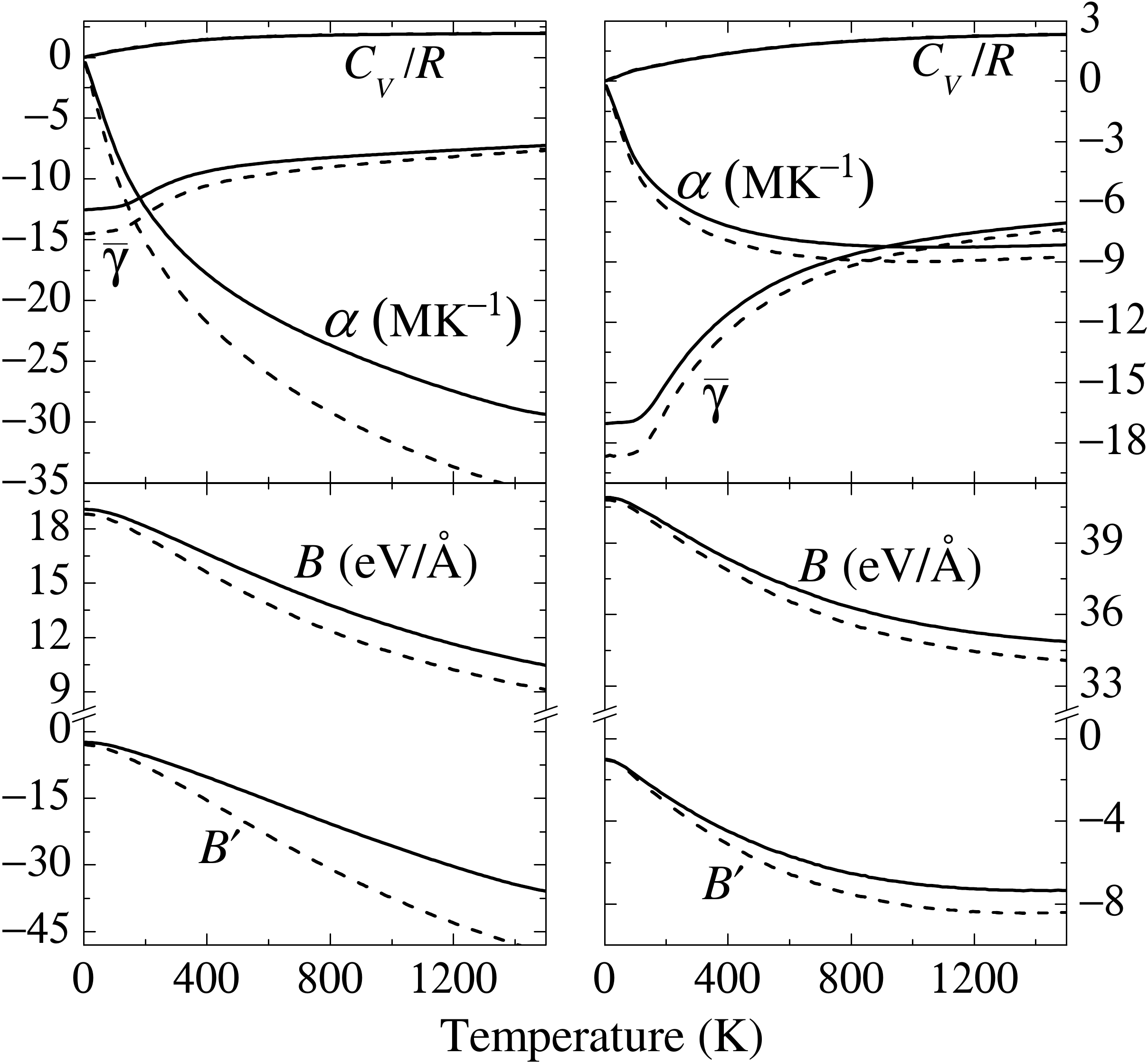}
\end{center}
\caption{\label{quasimodel} The temperature dependence of the coefficient of thermal expansion $\alpha$, overall Gr\"{u}neisen parameter $\overline{\gamma}$, bulk modulus $B$ and its pressure derivative $B^\prime$, calculated at two pressures (solid: $p=0$; dashed: $p=0.1$ eV/{\AA}) in a quasi-harmonic model interpretation of the diatomic ceramic (left) and triatomic hybrid models (right) examined in this paper. Note that $C_V/R$ of the hybrid model is less than 3 since the high-frequency optic branch is not significantly populated at these temperatures.}
\end{figure}

\section{Thermodynamics in the quasi-harmonic approximation}

The Helmholtz free energy $F$ of the two models was calculated in analytical form in the quasi-harmonic approximation \cite{Ashcroft 1976}, from which pressure, bulk modulus, the pressure derivative of bulk modulus, heat capacity, and the coefficient of thermal expansion were derived using standard thermodynamic relations: $p=-(\partial F/\partial L)_T$, $B=-L(\partial p/\partial L)_T$, $B^{\prime}=(\partial B/ \partial p)_T$, $C_V = -T(\partial^2 F/\partial T^2)_L$ and $\alpha=(1/L)(\partial L / \partial T)$ respectively. The quantities were computed self-consistently for two pressures and a range of temperatures, and are shown in Figure \ref{quasimodel}, together with the overall Gr\"{u}neisen parameter $\overline{\gamma}$ defined as the weighted average of the individual $\gamma_{s,q}$ values.

The results shown in the figures were obtained from self-consistent calculations using scripts written for MATLAB. The values of parameters of the models here were chosen in part to give phonon frequencies and values of physical properties comparable to known NTE materials, but also to ensure that the variations in frequencies can be seen on the same graphs. The force constants for the ceramic model were $k = 0.17$~eV\,{\AA}$^{-2}$ and  $K=J=0.0067$~eV\,{\AA}$^{-2}$; for the hybrid model $k=k^\prime=0.23$~eV\,{\AA}$^{-2}$, $J'=0.12$~eV\,{\AA}$^{-2}$, and $K=J=0.0067$~eV\,{\AA}$^{-2}$.  We set all mass values to $1$~g\,mol$^{-1}$; the effect of changing the masses can be reproduced by corresponding changes in the force constants. We set $K = J$ for convenience for plotting in Figures \ref{phonons} and \ref{quasimodel}; in reality, as discussed in the introduction, it is more likely that $K \gg J$, because $K$ will represent the energy associated with bending bonds within a quasi-rigid polyhedral group of atoms, and $J$ will represent the energy associated with bending the angles at the shared apex or via a shared ligand.

The standard Gr\"{u}neisen quasi-harmonic formulation gives $\alpha = \overline{\gamma} \,C_V / BL$. The value of $\overline{\gamma}$ becomes slightly less negative at higher temperatures (Figure \ref{quasimodel}) because of the contributions of the higher-frequency modes, which are more populated at higher temperatures but have smaller values of $\gamma_{s,q}$. The bulk modulus $B$ decreases on heating. Its temperature dependence in the high-temperature limit (when $\tau > \hbar \omega_\mathrm{max} \approx \hbar \sqrt{8K + 4J}$) can be derived from the analytical form of $F$ and $p$ as
\begin{equation}\label{bulk}
B \approx L\frac{{\partial ^2 \Phi }}{{\partial L^2 }} + \frac{\tau }{L}\overline \gamma
\end{equation}
where $\tau =k_\mathrm{B} T$, and  $\Phi  = \left( {k/2} \right)\left( {L - L_0 } \right)^2$ is the lattice energy. The temperature dependence of $B$ in Figure \ref{quasimodel} is mostly due to the factor of $\tau$. In turn, the temperature dependence of $B$ dominates the temperature dependence of $\alpha$ in the quasi-harmonic approximation, making it more negative at higher temperature. We discuss this later in this paper.

Figure \ref{quasimodel} shows that $\alpha$ also becomes more negative at higher pressure. This can be directly linked both to the larger negative value of $\overline{\gamma}$ and to the decrease of $B$ with pressure seen in Figure \ref{quasimodel}, and is consistent with the thermodynamic relation $(\partial \alpha/\partial p)_T = (1/B^2)(\partial B/\partial T)_p$.

The decrease of $B$ with pressure seen in Figure \ref{quasimodel}---the phenomenon called \textit{pressure-induced softening}~\cite{Pantea 2006,Chapman 2007,Fangexpression 2014}---is shown by the negative value of $B^\prime$. In the high-temperature limit we can write an equation for $B^\prime$:
\begin{equation}
\label{dbulk_highT}
B^\prime \approx  - \frac{L}{B}\left( {\frac{{\partial ^2 \Phi }}{{\partial L^2 }} + L\frac{{\partial ^3 \Phi }}{{\partial L^3 }}} \right) + \frac{\tau }{L}\frac{{\overline \gamma  }}{B} + \frac{{2\tau }}{L}\frac{{\partial \overline \gamma  }}{{\partial p}} + \frac{{\tau B}}{L}\frac{{\partial ^2 \overline \gamma  }}{{\partial p^2 }}
\end{equation}
The first term has a value of $-1$ in one dimension, with corresponding values of $0$ and $+1$ in two and three dimensions respectively from thermodynamic relations (see `Methods'). Writing the mode frequencies in approximate form $\omega^2(s,q) = a_{s,q}J + c_{s,q} k e+ O(e^2)$ and noting that the coefficients $a_{s,q}$ and $c_{s,q}$ always have positive values in both models, we find
\begin{eqnarray}
\label{gamma_derivative}
\gamma_{s,q} &=& - {c_{s,q} k }/{\omega^2(s,q)} < 0 \nonumber \\
{\partial \gamma_{s,q}}/{\partial p} &=& - {(c_{s,q} k)^2}/{\left(2B \omega^4(s,q)\right)} < 0  \nonumber \\
{\partial^2 \gamma_{s,q}}/{\partial p^2} &=& - {(c_{s,q} k)^3}/{\left(B^2 \omega^6(s,q)\right)} < 0 .
\end{eqnarray}
Thus, $\overline{\gamma}$ and its first and second pressure derivatives will also be negative. The contribution from these terms to the value of $B^\prime$ can prevail on heating as shown by Equation \ref{dbulk_highT}. This demonstrates that a negative $B^\prime$ is inevitable consequence of NTE in this model, which has been hypothesised but not previously justified theoretically~\cite{Fangzeolite 2013,Fangexpression 2014}. Furthermore, our model shows that $B^\prime$ must become more negative on compression due to the softening of both the vibration modes and the bulk modulus under pressure (Figures \ref{phonons} and \ref{quasimodel} and Equation~\ref{gamma_derivative}).

The preceding discussion around Equation \ref{dbulk_highT} also shows that the value of $B^\prime$ will become more negative on heating, and this is confirmed by Figure \ref{quasimodel}. For three-dimensional systems it might be expected that $B^\prime$ will have a positive value at zero temperature~\cite{Fangexp 2013}, with a slight reduction due to zero-point motions (although there is no constraint that this should be negative). But the vibrational contribution to $B^\prime$ on heating is negative and will become more dominant, eventually making $B^\prime$ negative, regardless of dimension. Thus we predict from our model that NTE materials will have a characteristic variation of $B^\prime$ with temperature, and this has in fact been observed in zeolites~\cite{Fangzeolite 2013,Fangexpression 2014} and Zn(CN)$_2$~\cite{Fangexp 2013}.

It is the case that in real three-dimensional systems there will be contributions from a large number of vibrational modes with positive Gr\"{u}neisen parameters. The analysis presented above will hold provided the NTE phonons are more strongly weighted in the overall Gr\"{u}neisen parameter $\overline{\gamma}$. This will be the case if the frequencies of the NTE phonons are low. In our model, this specifically means the lowest-frequency modes, whose values are determined by the value of $V_3$ (equation \ref{potential_3}).

\section{The role of anharmonic interactions}

Whilst our models show that both $\alpha$ and $B^\prime$ continue to decrease on increasing temperature, albeit changing less at high temperatures, this is actually different from what is normally seen in NTE materials, where both quantities become less negative on heating \cite{Li 2011,Mary1996,Evans1996,Sanson2006,Rodriguez2009,Attfield 1998,Fangzeolite 2013,Chapman 2005,Fangmd 2013,Zhou 2008, Fangexp 2013}. This indicates an insufficiency of the quasi-harmonic model as defined so far, and has to be tackled by including the effects of mode anharmonicity.

An anharmonic model was built by considering an explicit temperature dependence of phonon frequencies, using the standard approach for phonon--phonon interactions that underpins the soft-mode theory of displacive phase transitions\cite{Dove 1993}. This leads to a renormalisation of the phonon frequency:
\begin{align}\label{freq_anharm}
\widetilde{\omega}^2(s,q)&=a_{s,q}J + c_{s,q} ke + C_{s,q}T\sum\limits_{s^\prime,q^\prime}1/(a_{s^\prime,q^\prime} J + c_{s^\prime,q^\prime} ke) \nonumber \\
&= a_{s,q}J + c_{s,q} k \left[e (1-T/T_\mathrm{a}(s,q)) + (J/k)(T/T_\mathrm{a}(s,q))\right]
\end{align}
where $C_{s,q}>0$ is the coupling factor measuring the phonon-phonon interactions. The sum is over all modes except the one with zero frequency. We set $a_{s^\prime,q^\prime} = c_{s^\prime,q^\prime}$, which strictly holds for the important acoustic branch. We have defined $1/T_\mathrm{a}(s,q)  = C_{s,q}(\sum\limits_{s^\prime,q^\prime \ne 0} {1/c_{s^\prime,q^\prime}})/c_{s,q} J^{2}$ as a measure of the magnitude of the intrinsic anharmonicity; $T_\mathrm{a}(s,q)$ is the temperature beyond which the mode frequency starts to \emph{increase} on compression. Since the transverse acoustic modes at $q\neq0$ and $\neq\tfrac12$ are found to have smaller $c_{s,q}$ compared to the optic ones, these modes will have smaller $T_\mathrm{a}(s,q)$---hence anharmonic effects will be more important at any given temperature---compared to the optic modes in this model. This is consistent with what has been seen in real NTE materials~\cite{Mittal 2011,Fangmd 2013}. For simplicity we use a uniform value of $T_\mathrm{a}$ for all vibrational modes. Apart from added simplicity, this is expected to better simulate the case of a three-dimensional material, where there are more optic modes and hence more contribution to the overall anharmonicity from the optic modes than in the one-dimensional model, where the relative contribution to the anharmonicity from the acoustic modes is higher.

\begin{figure}[t]
\begin{center}
\includegraphics[width=8.0cm]{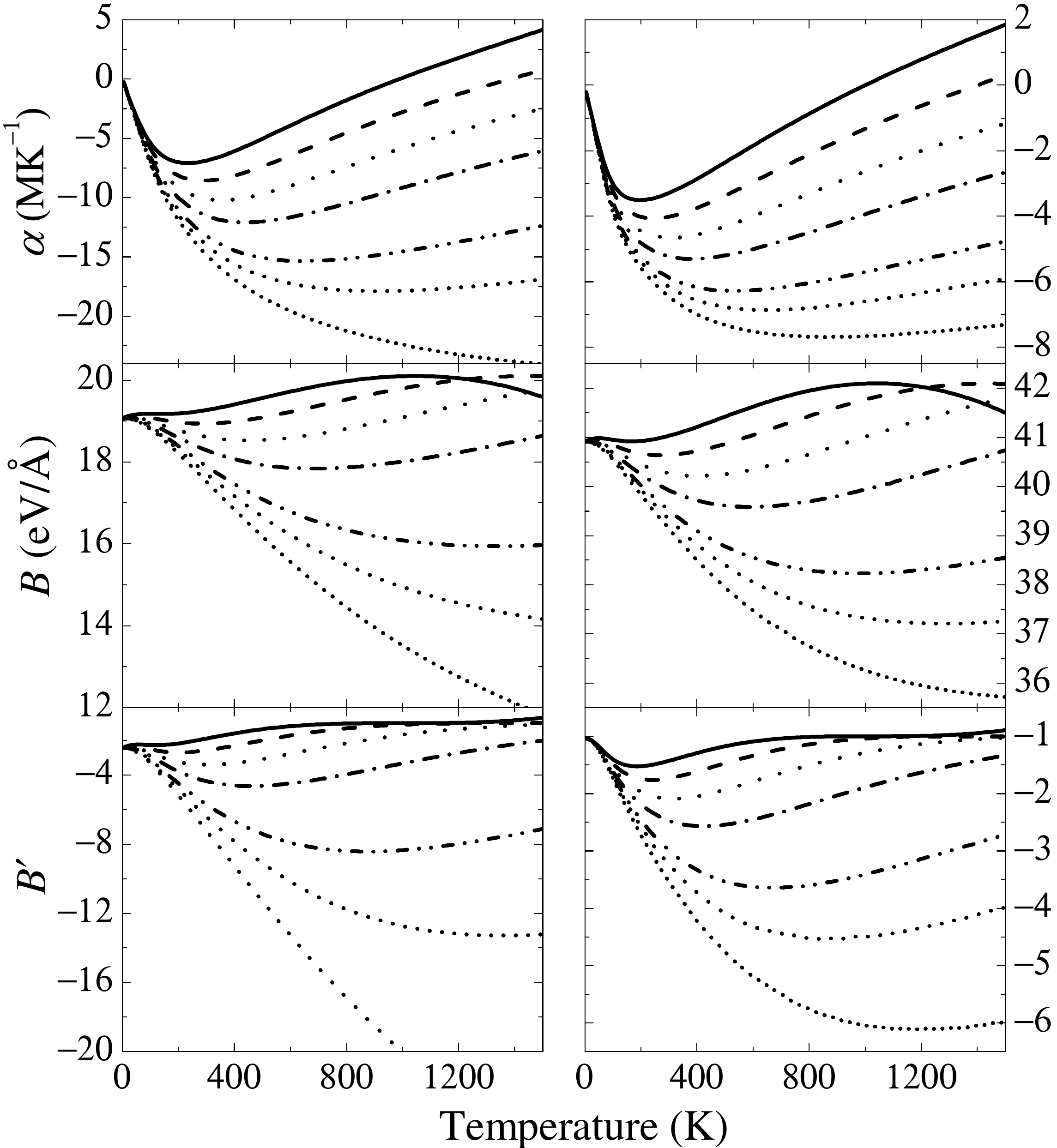} 
\end{center}
\caption{\label{anharmmodel} Temperature dependence of $\alpha$, $B$ and $B^\prime$ at zero pressure for the ceramic (left panel) and the hybrid (right panel) models. In each case curves decrease in anharmonicity from top to bottom (solid to short-dotted curves), with $T_\mathrm{a} = 1000, 1400, 2000, 3000, 6000, 10\,000, 30\,000$~K respectively, showing the trend to approach quasi-harmonic results from the anharmonic model. At $T = T_\mathrm{a}$, $\alpha \approx 0$, and for $T>T_\mathrm{a}$ the models show positive thermal expansion.}
\end{figure}

Figure~\ref{anharmmodel} shows $\alpha$, $B$, and $B^\prime$ recalculated for both ceramic and hybrid models including the anharmonic terms with different values of $T_\mathrm{a}$. The temperature dependence of $\alpha$ is determined by the fact that the renormalised mode frequencies (Equation~\ref{freq_anharm}) are stiffened on heating, leading to renormalised values of mode Gr\"{u}neisen parameters
\begin{align}\label{renormgrun}
\widetilde{\gamma}_{s,q}=- c_{s,q} k (1-T/T_\mathrm{a}) / \widetilde{\omega}^2(s,q).
\end{align}

Thus, according to Equations~\ref{freq_anharm} and~\ref{renormgrun}, at $T=T_\mathrm{a}$, $\widetilde{\omega}$ becomes independent of $e$ and $\widetilde{\gamma}_{s,q}$ becomes zero, leading to zero overall Gr\"{u}neisen parameter hence zero $\alpha$. When $T>T_\mathrm{a}$, the coefficient of $e$ in the renormalised frequency becomes negative and $\widetilde{\gamma}_{s,q}$ is positive, resulting in positive overall Gr\"{u}neisen parameter hence positive $\alpha$.

In addition to anharmonicity, the other reason for the reduction of NTE on heating is the population of high-energy modes with positive mode Gr\"{u}neisen parameters. The relative contributions of these two effects can be estimated by the extent to which quasiharmonic calculations of $\alpha(T)$ from lattice dynamics calculations reproduce the experimental data. On this basis, by comparing to the curves of different cases of anharmonicity in Figure~\ref{anharmmodel}, we found that ReO$_3$~\cite{Chatterji 2008}, Cu$_2$O~\cite{Bohnen 2009}, and ScF$_3$ \cite{Li 2011} correspond to cases of strong anharmonic effects, while examples of weaker anharmonicity are Zn(CN)$_2$ \cite{Chapman 2005,Fangmd 2013} and MOF-5 \cite{Zhou 2008}. These examples empirically suggest that anharmonicity is likely to be more important in ceramic systems than MOF systems in general: indeed, since MOFs will have more atoms in the unit cell and hence more phonon modes than ceramics, it might be expected that the higher-frequency modes are more likely to be significantly populated in these systems, compared to ceramics, at temperatures where anharmonicity remains negligible.

In the recently reported case of Ag$_2$O, $\alpha(T)$ acts like the strong anharmonic case in Figure~\ref{anharmmodel} but starts to become more negative at temperatures $>250$ K~\cite{Gupta2012}. This behaviour suggests a negative coupling factor $C_{s,q}$, hence negative $T_\mathrm{a}(s,q)$ for some high-energy modes in Equation~\ref{freq_anharm}. According to Equation~\ref{renormgrun}, with elevated temperature, $\widetilde{\gamma}_{s,q}$ of these modes will become more negative and will contribute more to the overall Gr\"{u}neisen parameter when the modes are more occupied. This will in turn begin to enhance the NTE at relatively high temperatures, exactly as seen in the temperature-dependent behaviour of $\alpha(T)$ in Ag$_2$O~\cite{Gupta2012,Lan2014}.

With high anharmonicity the value of $B$ will initially decrease on heating but then quickly start to increase, only decreasing again when $T>T_\mathrm{a}$ (Figure \ref{anharmmodel}). The prediction of the counterintuitive increase of $B$ upon heating which corresponds to warm hardening of the structure has recently been demonstrated in Sc$_{1-x}$Y$_x$F$_3$\cite{Morelock 2013}. For weaker anharmonicity, $B$ will only decrease on heating as for the quasi-harmonic model.

Comparing Figures \ref{quasimodel} and \ref{anharmmodel} it can be seen that the effect of anharmonicity is to increase (\emph{i.e.}, make less negative) the value of $B^\prime$ at higher temperature. For weaker anharmonicity, the value of $B^\prime$ will reach a minimum at higher $T$, and thereafter tend towards zero on heating. This effect has been seen in cubic NTE zeolites \cite{Fangzeolite 2013} and Zn(CN)$_2$ \cite{Fangexp 2013}. On the other hand, with very large anharmonicity $B^\prime$ increases with temperature beyond $T_\mathrm{a}$, as seen in materials with positive thermal expansion~\cite{Zhang2007,Song2012,Sun2013}, and in a three-dimensional system will not have a negative value. This has been observed in ScF$_3$ \cite{Morelock 2013}, where $\alpha$ becomes positive around $1100$ K \cite{Li 2011}, and corresponds to the case shown as the solid line in the left panel of Figure \ref{anharmmodel}.

\section{Conclusions}

In summary, we have investigated simple models designed to mimic the essential features of many ceramic and hybrid NTE systems, particularly taking account of the role of transverse vibrations of bridging atoms and recognising that often bridging atoms are part of a stiff polyhedral group of atoms. We have shown that pressure-induced softening will in many cases be an inevitable consequence of NTE. Moreover, the model shows how  NTE, elastic stiffness and pressure-induced softening will vary with both temperature and pressure.

We have also shown how the basic picture is modified by intrinsic anharmonic interactions beyond those captured in the quasiharmonic model of NTE. In particular, both the NTE and pressure-induced softening are weakened at high temperature due to anharmonic interactions, and the same interactions lead to warm hardening.

The objective of studying these models has been to provide a framework in which both NTE and pressure-induced softening can be understood as arising from a common mechanical origin, something that has previously been hinted at but not explicitly demonstrated. We hope that these models will prompt further experimental investigation of the links between these properties in real materials.

\section{Acknowledgements}

We gratefully acknowledge financial support from the CISS of Cambridge Overseas Trust (HF). We thank the support of the CamGrid high-throughput environment of the University of Cambridge and the UK HPC Materials Chemistry Consortium, funded by EPSRC (EP/F067496).

\end{document}